\documentclass[amsmath,amssymb,preprint,floatfix]{revtex4-1}

\usepackage{hyperref}
\usepackage{epsfig}
\usepackage{graphicx}
\usepackage{subfigure}
\usepackage{latexsym}
\usepackage{color}
\usepackage{fullpage}
\usepackage{dcolumn}
\usepackage{bm}
\usepackage{ulem}
\usepackage{units}

\begin{document}

\title{Perpendicular magnetic anisotropy of cobalt films intercalated under graphene}%

\author{N. Rougemaille$^{1,2}$, A. T. N'Diaye$^{3}$, J. Coraux$^{1,2}$, C. Vo-Van$^{1,2}$, O. Fruchart$^{1,2}$, A. K. Schmid$^{3}$}

\address{$^1$ CNRS, Inst NEEL, F-38042 Grenoble, France\\$^2$ Univ. Grenoble Alpes, Inst NEEL, F-38042 Grenoble, France\\$^3$ National Center for Electron Microscopy, Lawrence Berkeley National Laboratory, One Cyclotron Road, Berkeley, California 94720, USA}

\date{\today}%

\begin{abstract}
Magnetic properties of nanometer thick Co films intercalated at the graphene/Ir(111) interface are investigated using spin-polarized low-energy electron microscopy (SPLEEM) and Auger electron spectroscopy. We show that the graphene top layer promotes perpendicular magnetic anisotropy in the Co film underneath, even for relatively thick intercalated deposits. The magnetic anisotropy energy is significantly larger for the graphene/Co interface than for the free Co surface. Hybridization of the graphene and Co electron orbitals is believed to be at the origin of the observed perpendicular magnetic anisotropy.   
\end{abstract}


\maketitle

Owing to its peculiar electronic band structure, high charge carrier mobility and long spin diffusion length, graphene is a promising two-dimensional material for microelectronics and spintronics. Exciting spin-dependent effects have been observed and predicted. Gate tunable spin transport in non-local spin valve devices,\cite{Han2011} tunnel spin injection,\cite{Tombros2007} spin-filtering,\cite{Son2006,Karpan2007} and large magnetoresistance\cite{Karpan2007} are a few examples of fascinating phenomena. Graphene also shows interesting magnetic properties when in contact with a ferromagnetic metal (FM). For instance, graphene carries a net magnetic moment when deposited on Ni(111) or Fe/Ni(111),\cite{Weser2011} and a 13\,meV spin splitting can be induced in graphene due to proximity with a heavy element.\cite{Varykhalov2008}

While these reports illustrate potential uses for graphene integration within magnetic device structures, the influence of graphene on the magnetic properties of a FM is still largely unexplored. In particular, interfaces with non-magnetic overlayers (adsorbates, capping layers) generally affect the magnetic anisotropy energy (MAE) of thin layers, \cite{Elgabaly2008,Santos2012} and it is interesting to explore how an interface with graphene would influence the MAE of a thin FM film. However, the fabrication of atomically flat graphene/FM and FM/graphene interfaces poses challenges. The synthesis of well-ordered epitaxial graphene on metal surfaces requires processing at high-temperature, where thin film dewetting or interfacial intermixing might occur. On the other hand, metal evaporation on graphene often yields clustered deposits.\cite{Binns1999,Ndiaye2009} Alternative routes have been proposed to fabricate atomically flat interfaces, such as metal deposition on graphene using pulsed laser deposition,\cite{VoVan2010} and metal intercalation between graphene and its substrate.\cite{Tontegode1991} Here we use an intercalation process to prepare homogeneous epitaxial cobalt layers, sandwiched between an Ir(111) substrate and a monolayer of graphene. 

\begin{figure}
  \begin{center}
  \includegraphics[width=80mm]{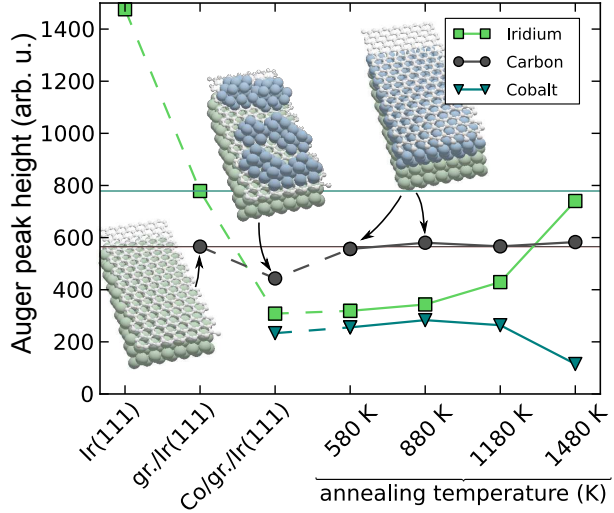}
  \caption{\label{Figure1} (color online) Intercalating Co grown on graphene/Ir(111). Tracking elemental Auger peak intensities during graphene and Co deposition, Co intercalation (annealing in the range of 580-880\,K) and loss of Co during over-annealing above ~880\,K.} 
  \end{center}
\end{figure}

Experiments were conducted in the SPLEEM\cite{Bauer2005-1,Bauer2005-2} instrument of the National Center for Electron Microscopy of the Lawrence Berkeley Lab\cite{Rougemaille2010}. Samples were grown in situ, in ultra-high vacuum conditions (base pressure in the low 10$^{-10}$\,mbar or below). An Ir(111) single crystal is used as a substrate and cleaned with repeated cycles of Ar sputtering and high temperature flash (1200$^\circ$C) under oxygen ($10^{-8}$\,mbar at 800$^\circ$C). Graphene was prepared on freshly cleaned Ir(111) surfaces by chemical vapor deposition with ethylene (pressure ranging from $10^{-8}$ to $10^{-7}$\,mbar) at 900\,K substrate temperature. At this temperature, nucleation density is high, typically hundreds of graphene islands per $\mu m^{2}$.\cite{Coraux2009} After cooling the graphene/Ir(111) substrate to room temperature, Co was deposited from electron-beam evaporation sources at rates between 0.5 to 3\,monolayer (ML) per minute.\cite{noteonML} Finally, annealing to a temperature in the range of 500-600\,K causes an intercalation process in which Co and graphene exchange places, resulting in a structure where the Co film is sandwiched between the Ir(111) substrate and the graphene monolayer. Fig.\ref{Figure1} summarizes element specific peak heights from Auger-electron spectra collected during the preparation of a graphene/Co(2 ML)/Ir(111) sample. One monolayer of graphene covering the clean Ir(111) substrate produces a 272 eV carbon peak at the level marked by a black line and reduces the size of the 39 eV iridium peak to the level marked by a green line. Subsequent room temperature deposition of a dose of cobalt equivalent to 2 ML thickness reduces the C and Ir peaks. However, even moderate annealing, holding 580 +/- 20\,K for 5 minutes, already brings the carbon signal back to the level which corresponds to bare graphene at the top of the sample (black line). After subsequent annealing steps up to 880\,K (always holding target temperatures within +/- 20\,K for 5 minutes) the Auger peak heights remain constant until, at much higher annealing temperature in the range above ~880\,K, the gradual increase of the Ir signal in conjunction with reduction of the Co signal indicates that integrity of the cobalt layer is lost. Co-Ir is an isomorphic binary alloy system and Co might be dissolving into the Ir crystal (and/or might be lost due to sublimation). However, stability of the three elemental Auger peaks within a wide range of annealing temperature between about 580-880\,K indicates that the Co/Ir(111) interfaces are kinetically quite stable. 
Note that in this experiment a thin Co deposit was used (equivalent to about 2 ML) so that all three elemental Auger electron peaks can be tracked in all phases of the sample preparation. In all other experiments described in the rest of this letter, Co deposits were thicker (10 ML or more - in that case the 39 eV iridium peak and the 272 eV carbon peak from buried graphene are essentially suppressed). 

\begin{figure}
  \begin{center}
  \includegraphics[width=80mm]{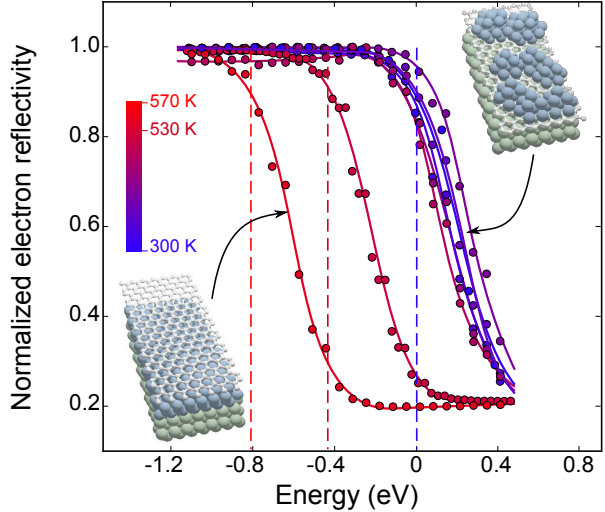}
  \caption{\label{Figure2} (color online) Intercalating Co grown on graphene/Ir(111). Tracking normalized electron reflectivity by SPLEEM, as a function of the incident energy of the electron beam for different annealing temperatures. Before annealing, a graphene/Ir(111) film was buried under a Co film, which has a high work function. During annealing in the temperature range above ~530\,K progressive reduction of the surface work function indicates formation of graphene termination at the sample surface, as the Co layer intercalates between the graphene and the Ir(111) substrate.}
  \end{center}
\end{figure}

To track the intercalation process in case of larger Co deposits we use in-situ low-energy electron microscopy observation, where the presence of graphene at sample surfaces can generate a strong signal. Low energy electron microscopes are very sensitive tools to resolve the value of the surface work function with high spatial- and energy resolution, \cite{unal2009} and graphene has a characteristically low work function compared to the metal surfaces this work is concerned with. \cite{Tontegode1991,Loginova2009} When the energy of the incident beam is lower than the surface work function, electron reflectivity of the surface is essentially 100 \%, while reflectivity decreases by a substantial factor when electrons have enough kinetic energy to enter the crystal. This is shown in Fig.~\ref{Figure2}. As long as a freshly prepared Co/graphene/Ir(111) sample has not been annealed, the incident electron energy threshold of 100\% reflectivity remains high (we reference to this energy value as 0 eV in the plot). As soon as annealing temperature is sufficiently high to trigger intercalation, above ~530\,K, the emergence of graphene at the sample surface induces a drop of this energy threshold by a good fraction of one eV. In LEEM images acquired under these conditions, metal-covered samples appear very bright and graphene terminated structures appear much darker.

In all cases of intercalated graphene/Co/Ir(111) structures prepared in this work (we made several dozen samples) the in-situ LEEM observations never show dewetting or diffusive aggregation of large Co islands or clusters. This is interesting because in the {\it absence} of graphene, we observe the formation of 3D islands (or clusters) between 800\,K and 900\,K. Such processes are common and clearly visible in in-situ LEEM images, as has been also documented in many similar metal-on-metal systems including cobalt \cite{Ding2005} or copper \cite{Ling2004} on Ru(0001), chromium \cite{Mccarty2009} and iron \cite{Rougemaille2006} on W(011), etc. Given the prevalence of 3D islanding in metal-on-metal systems and the efficiency of LEEM in detecting 3D islanding phenomena, the apparent suppression of 3D islanding by the presence of a layer of graphene is interesting. One might speculate that the graphene layer stabilizes homogeneous and flat intercalated films by acting as a surfactant; however, understanding the underlying kinetics and energetics merits additional work beyond the scope of this letter. 
    
\begin{figure}
  \begin{center}
  \includegraphics[width=80mm]{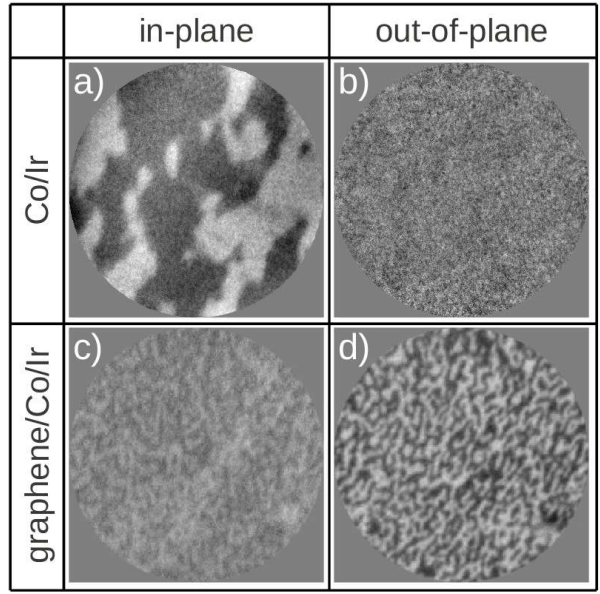}
  \caption{\label{Figure3} 10\,$\mu$m field of view SPLEEM images of 11 ML of Co deposited on Ir(111) (a-b) and intercalated at the graphene/Ir(111) interface (c-d). Left and right images show in-plane and out-of-plane components of magnetization, respectively.}
  \end{center}
\end{figure}

Here we focus our attention on the magnetic properties of the Co layers. The example shown in Fig.~\ref{Figure3} compares magnetization direction and magnetic domain structure of 11\,ML Co films grown directly on Ir(111) (Fig.~\ref{Figure3}a-b) and an intercalated graphene/Co/Ir(111) structure with equal Co thickness (Fig.~\ref{Figure3}c-d). Both samples were annealed to 570\,K. While the 11\,ML Co film on Ir(111) has in-plane magnetization (no magnetic contrast is observed when probing the out-of-plane component of magnetization, Fig.~\ref{Figure3}b), an 11\,ML Co film intercalated at the graphene/Ir(111) interface is out-of-plane magnetized. Note however that a small in-plane contrast is also observed, indicating that the magnetization is not purely out-of-plane at this Co thickness, but slightly canted. For Co/Ir(111), we observe a thickness-dependent spin reorientation transition and find that the transition between out-of-plane and in-plane magnetization occurs at about $t_{1}\approx\unit[6]{Co\,ML}$. In the case of graphene/Co/Ir(111), this transition occurs for thicknesses $t_2$ ranging between 12 and 13 Co ML \textit{i.e.} approximately twice the value found for vacuum/Co/Ir(111). This shows that adding a graphene/Co interface at the top of an 11\,ML Co/Ir(111) film enhances the PMA in the Co layer.

To estimate the contribution of the graphene/Co interface to the PMA, we start by assuming that the MAE is uniaxial with second order. The energy density of magnetic layers can then be written $E=-K\cos^2\theta$, where $\theta$ is the angle between the magnetization and the normal to the film and the total magnetic anisotropy, $K$ includes shape anisotropy $K_\mathrm{S}=(1/2)\mu_0M_\mathrm{S}^2$ and magnetocrystalline anisotropy $K_\mathrm{mc}$. The latter may be split into the volume contribution $K_\mathrm{V}$ and surface/interface contributions: $K_\mathrm{mc}=K_\mathrm{V}+(K_{\mathrm{Co/Ir}}+K_{\mathrm{vacuum/Co}})/t$ for bare Co/Ir(111) films and $K_\mathrm{mc}=K_\mathrm{V}+(K_{\mathrm{Co/Ir}}+K_{\mathrm{Gr/Co}})/t$ for graphene covered films, with $t$ the film thickness. Critical thicknesses of bare and graphene covered films, $t_{1}$ and $t_{2}$, correspond to the condition $K_\mathrm{mc}=K_\mathrm{S}$. In our case the shape anisotropy is constant, so we can write 
$K_\mathrm{V}+(K_{\mathrm{Co/Ir}}+K_{\mathrm{vacuum/Co}})/t_{1}=K_\mathrm{V}+(K_{\mathrm{Co/Ir}}+K_{\mathrm{Gr/Co}})/t_{2}$, which permits us to estimate the contribution of the graphene/Co interface to the PMA: 
$K_{\mathrm{Gr/Co}}=(t_{2}/t_{1}-1)K_{\mathrm{Co/Ir}}+(t_{2}/t_{1})K_{\mathrm{vacuum/Co}}$.
The interfacial energy $K_\mathrm{Co/Ir}$ was reported to be $\unit[0.8]{mJ/m^2}$.\cite{Broeder1991} While experiments \cite{Rusponi2003} showed that the value of $K_{\mathrm{vacuum/Co}}$ is close to zero in the case of Co films strained to match the 0.277 nm lattice constant of Pt(111), lattice strain-resolved ab-initio calculations \cite{ElGabaly2006} showed that the value of the surface anisotropy increases by approximately $\unit[0.3]{mJ/m^2}$ when the in-plane lattice constant is reduced to the value of bulk Co, 0.251 nm.
We observed the critical thickness to increase by the factor 13/6 when the Co films are intercalated under graphene. Assuming this increase is driven by the difference of interfacial anisotropy between Co-vacuum and Co-graphene interfaces alone, it follows that the value of $K_\mathrm{Gr/Co}$ is approximately $\unit[1.6]{mJ/m^2}$. This value is relatively large compared to values usually reported for Co/metal interfaces, possibly suggesting that additional mechanisms are at work in our system. For example, we can not exclude that the intercalation process and the surfactant role of the graphene layer affect the wetting of the Co film on the Ir(111) surface and its structural properties. These effects might induce strain in the FM and induce changes in the crystallography, thus modifying the magnetic anisotropy. 

In conclusion, we use spin-polarized low-energy electron microscopy and Auger electron spectroscopy to study the intercalation of cobalt deposited on top of graphene/Ir(111), and we find that a graphene top layer affects the magnetic properties of nm-thick Co films on Ir(111). We show that in the intercalated cobalt, perpendicular magnetic anisotropy is favored over an unusually large thickness range. Compared to the vacuum/Co interface, the MAE is significantly larger for a graphene-terminated Co surface. The hybridization of the Co and graphene electron orbitals very likely play a key role in this unusual MAE. These result open new perspectives for graphene-based spintronic devices with perpendicular magnetic anisotropy.

C.V.V acknowledges financial support from the Fondation Nanosciences. A.T.N. acknowledges the Alexander von Humboldt Foundation for a Feodor Lynen research fellowship. This work was partially supported by the French ANR contract ANR-2010-BLAN-1019-NMGEM. Experiments were performed at the National Center for Electron Microscopy, Lawrence Berkeley National Laboratory, supported by the Office of Science, Office of Basic Energy Sciences, Scientific User Facilities Division, of the U.S. Department of Energy under Contract No. DE-AC02-05CH11231.


\begin{thebibliography}{2}

\bibitem{Han2011}
W. Han and R. K. Kawakami, Phys. Rev. Lett. {\bf 107}, 047207 (2011)

\bibitem{Tombros2007}
N. Tombros, C. Jozsa, M. Popinciuc, H. T. Jonkman, and B. J. van Wees, Nature {\bf 448}, 571 (2007)

\bibitem{Son2006}
Y. W. Son, M. L. Cohen, and S. G. Louie, Nature {\bf 444}, 347 (2006)

\bibitem{Karpan2007}
V. M. Karpan, G. Giovannetti, P. A. Khomyakov, M. Talanana, A. A. Starikov, M. Zwierzycki, J. van den Brink, G. Brocks, and P. J. Kelly, Phys. Rev. Lett. {\bf 99}, 176602 (2007)

\bibitem{Weser2011}
M. Weser, E. N. Voloshina, K. Horna, and Yu. S. Dedkov, Phys. Chem. Chem. Phys. {\bf 13}, 7534 (2011)

\bibitem{Varykhalov2008}
A. Varykhalov, J. S\'{a}nchez-Barriga, A. M. Shikin, C. Biswas, E. Vescovo, A. Rybkin, D. Marchenko, and O. Rader, Phys. Rev. Lett. {\bf 101}, 157601 (2008)

\bibitem{Elgabaly2008}
F. El Gabaly, K. F. McCarty, A. K. Schmid, J. de la Figuera, M. C. Munoz, L. Szunyogh, P. Weinberger, S. Gallego, New J. of Physics {\bf 10}, 073024 (2008)

\bibitem{Santos2012}
B. Santos, S. Gallego, A. Mascaraque, K. F. McCarty, A. QuesadaA. T. N'Diaye, A. K. Schmid, J. de la Figuera, Phys. Rev. B {\bf 85} 134409 (2012)

\bibitem{Binns1999}
C. Binns, S. H. Baker, C. Demangeat, and J. C. Parlebas, Surf. Sci. Rep. {\bf 34}, 107 (1999)

\bibitem{Ndiaye2009}
A. T. N'Diaye, T. Gerber, M. Busse, J. Myslivecek, J. Coraux, and T. Michely, New J. Phys. {\bf 11}, 103045 (2009)

\bibitem{VoVan2010}
C. Vo-Van, Z. Kassir-Bodon, H. Yang, J. Coraux, J. Vogel, S. Pizzini, P. Bayle-Guillemaud, M. Chshiev, L. Ranno, V. Guisset, P. David, V. Salvador, and O. Fruchart, New J. Phys. {\bf 12}, 103040 (2010)

\bibitem{Tontegode1991}
A. Y. Tontegode, Progr. Surf. Sci. {\bf 38}, 201 (1991)

\bibitem{Bauer2005-1}
E. Bauer, in Magnetic Microscopy of Nanostructures, edited by H. Hopster, H.P. Oepen (Springer, Berlin), p.111 (2005)

\bibitem{Bauer2005-2}
E. Bauer, in Modern Techniques for Characterizing Magnetic Materials, edited by Y. Zhu (Kluwer, Boston), p.361 (2005)

\bibitem{Rougemaille2010}
N. Rougemaille and A. K. Schmid, Eur. Phys. J. Appl. Phys. {\bf 50}, 20101 (2010)

\bibitem{Coraux2009}
J. Coraux, A. {T. N'Diaye}, M. Engler, C. Busse, D. Wall, N. Buckanie, F.-J {Meyer zu Heringdorf}, R. {van Gastel}, B. Poelsema, and T. Michely, New J. Phys. {\bf 11}, 023006 (2009)

\bibitem{noteonML}
A monolayer is defined as the first atomic layer of Co that grows pseudomorphically on Ir(111).

\bibitem{unal2009}
B. Unal, Y. Sato, K. F. McCarty, N. C. Bartelt, T. Duden, C. J. Jenks, A. K. Schmid, P. A. Thiel, 
JVST A {\bf 27}, 1249 (2009)

\bibitem{Loginova2009}
E. Loginova, S. Nie, K. Th\"{u}rmer, N. C. Bartelt, and K. F. McCarty, Phys. Rev. B {\bf 80}, 085430 (2009)

\bibitem{Ding2005}
H. F. Ding, A. K. Schmid, D. Q. Li, K. Y. Guslienko, S. D. Bader, Phys. Rev. Lett. {\bf 94}, 157202 (2005)

\bibitem{Ling2004}
W. L. Ling, T.  Giessel, K. Thurmer, R. Q. Hwang, N. C. Bartelt, K. F. McCarty, Surface Science {\bf 570} (2004)

\bibitem{Mccarty2009} 
K. F. McCarty, J. C. Hamilton, Y. Sato, A. Saá, R. Stumpf, J. de la Figuera, K. Thurmer, F. Jones, A. K. Schmid, A. A. Talin, and N. C. Bartelt, New J. Phys. {\bf 11}, 043001 (2009)

\bibitem{Rougemaille2006}
N. Rougemaille, A. K. Schmid, J. Appl. Phys. {\bf 99}, 08S502 (2006)

\bibitem{Broeder1991}
F. J. A. den Broeder, W. Hoving, and P. J. H. Bloemen, J. Mag. Mag. Mater. {\bf 93}, 562 (1991)

\bibitem{Rusponi2003}
S. Rusponi, T. Cren, N. Weiss, M. Epple, L. Claude, P. Buluschek, and H. Brune, Nat. Mat., {\bf 2}, 546 (2003)

\bibitem{ElGabaly2006}
F. El Gabaly, S. Gallego, C. Mu\~{n}oz, L. Szunyogh, P. Weinberger, C. Klein, A. K. Schmid, K. F. McCarty, and J. de la Figuera, Phys. Rev. Lett. {\bf 96},  147202 (2006)

\end{thebibliography}
\end{document}